# Modeling and Simulation of Nitrogen Generation by Pressure Swing Adsorption for Power-to-Ammonia

Marcus J. Schytt[a], Lorenz T. Biegler[b], and John B. Jørgensen[a]*

[a] Technical University of Denmark, DTU Compute, Kongens Lyngby, Capital Region, Denmark
[b] Carnegie Mellon University, Department of Chemical Engineering, Pittsburgh, Pennsylvania, USA
* Corresponding Author: jbjo@dtu.dk.

## ABSTRACT

Power-to-ammonia (P2A) provides a carbon-free alternative to conventional ammonia production by replacing fossil-based feedstocks with electrolytic hydrogen and nitrogen from air separation. For decentralized P2A systems, pressure swing adsorption (PSA) offers a flexible alternative to cryogenic air separation. However, its industrial implementations are largely proprietary, and open, first-principles models capable of simulating its cyclic, nonlinear transport are scarce in literature. This work presents a first-principles, dynamic, one-dimensional model of a PSA superstructure for nitrogen generation, formulated with thermodynamically consistent equations of state, coupling multicomponent mass, energy, and momentum balances with kinetically limited adsorption on carbon molecular sieves. The resulting system of partial differential-algebraic equations (PDAEs) is semi-discretized using the finite volume method, integrated using diagonally implicit Runge-Kutta methods, and cyclic steady states (CSS) are computed via shooting-based solution methods. The framework is implemented in Julia, combining analytical derivatives with automatic differentiation and utilizing sparse linear algebra for efficient solution of the arising large nonlinear systems. The framework is demonstrated on a two-bed PSA cycle for air separation, comparing spatial and temporal discretization strategies, CSS solution methods, and the effects of ideal versus real-gas thermodynamics on predicted nitrogen purity and recovery. The proposed framework establishes an extensible basis for PSA simulation and optimization.

## INTRODUCTION

Conventional ammonia production relies on the famous Haber–Bosch process, combining nitrogen with hydrogen derived from fossil-based syngas. The resulting grey ammonia, of which approximately 80% is used for fertilizer production, accounts for about 4.5% of global industrial $CO_2$ emissions [1]. To mitigate this environmental impact, power-to-ammonia (P2A) has emerged as a carbon-free alternative, replacing fossil-based hydrogen with hydrogen produced by water electrolysis and sourcing nitrogen from atmospheric air. When powered by renewable electricity, P2A enables the production of green ammonia, which can serve both as a carbon-free hydrogen carrier and as a potential maritime fuel.

Nitrogen is traditionally produced by large-scale cryogenic air separation. However, for the smaller and decentralized production units envisioned in P2A systems, pressure swing adsorption (PSA) offers a flexible alternative. PSA achieves high-purity (95%+) nitrogen through cyclic adsorption and desorption of air components on carbon molecular sieves (CMS). Despite its widespread industrial adoption, PSA process design remains largely proprietary, and the open literature, often relying on commercial simulators, offers limited first-principles models capable of simulating its inherently cyclic operation. This limitation is exacerbated by strong multiscale coupling between transport, adsorption kinetics, and thermodynamics, making systematic computation of cyclic steady states (CSS) especially challenging.

Motivated by these gaps, this work develops a first-principles adsorption column model and superstructure for PSA-based nitrogen generation, which, using well-established numerical discretization methods, enables systematic computation of thermodynamically consistent cyclic operation.

## ADSORPTION COLUMN MODEL

We model adsorption columns with fixed beds composed of carbon molecular sieve pellets, specifically the commercial Takeda 3A CMS adsorbent. Parameters describing the adsorbent are given in the digital supplementary material. We model the adsorption column as a cylinder with length $L$ on the axial interval domain $\Omega = [0, L]$. We study a set of components $\mathcal{C} = \{N_2, O_2, Ar\}$ representative of the composition of air (78%, 21%, 1%). To model the adsorption process, the column is treated heterogeneously and divided into a mobile phase and a stationary phase with porosity $\varepsilon$. Accordingly, the column volume $V$ holds the gaseous volume $V_{\text{gas}} = \varepsilon V$, and the ratio between the gaseous and stationary volumes is given by $\varphi = \varepsilon/(1-\varepsilon)$. We denote mobile-phase concentrations by $c = [c_\alpha]_{\alpha \in \mathcal{C}}$, and stationary-phase concentrations by $q = [q_\alpha]_{\alpha \in \mathcal{C}}$. Moreover, we let $u$ denote the pseudo-homogeneous total internal energy density, accounting for both the mobile and stationary phases. The mass and energy balances of the adsorption column are then summarized by a system of one-dimensional transport equations subject to thermodynamic constraints

$$\partial_t c = -\partial_z N + R, \quad (1a)$$
$$\partial_t u = -\partial_z E + Q. \quad (1b)$$

Here, $N = [N_\alpha]_{\alpha \in \mathcal{C}}$ denotes the molar fluxes, $R = [R_\alpha]_{\alpha \in \mathcal{C}}$ the adsorption rates, $E$ the heat flux, and $Q$ the heat transfer rate.

### Thermodynamics

We introduce thermodynamic functions to evaluate bulk gas-phase properties. Assuming local thermodynamic equilibrium, the thermodynamic state is characterized by the local temperature, pressure, and composition of the system. While the composition is distributed between the mobile and stationary phases, the temperature and pressure are common to both phases. A gas phase with temperature $T$, pressure $P$, and moles $n = [n_\alpha]_{\alpha \in \mathcal{C}}$, can be described by thermodynamic functions

- Volume: $\mathcal{V}(T, P, n)$
- Enthalpy: $\mathcal{H}(T, P, n)$
- Internal energy: $\mathcal{U}(T, P, n)$

provided by an equation of state (EOS). The thermodynamic functions satisfy standard identities. In particular, they are related by

$$\mathcal{U}(T, P, n) = \mathcal{H}(T, P, n) - P\,\mathcal{V}(T, P, n). \quad (2)$$

Moreover, the thermodynamic functions are first-order homogeneous in the component moles $n$, i.e.,

$$\mathcal{V}(T, P, \lambda n) = \lambda \mathcal{V}(T, P, n), \quad \lambda > 0, \quad (3)$$

with analogous relations holding for $\mathcal{H}$ and $\mathcal{U}$. The EOS may be specified by ideal-gas correlations, with or without cubic real-gas modifications.

### Transport

The molar fluxes account for mobile-phase advection and are given by

$$N = vc, \quad (4)$$

assuming negligible diffusive contributions. Here, $v$ denotes the interstitial axial velocity. The velocity is related to the pressure gradient through a momentum balance. We employ Ergun's equation

$$-\partial_z P = 150\mu \left(\frac{\varphi}{d_p}\right)^2 v + 1.75\rho_{\text{gas}} \frac{\varphi}{d_p} v|v|. \quad (5)$$

Here, $\mu$ denotes the mobile-phase dynamic viscosity, $d_p$ the effective particle diameter, and

$$\rho_{\text{gas}} = \sum_{\alpha \in \mathcal{C}} c_\alpha M_\alpha, \quad (6)$$

the mobile-phase density, which is defined by molar masses $M = [M_\alpha]_{\alpha \in \mathcal{C}}$. We fix the dynamic viscosity to be that of air at ambient conditions and invert (5) to yield an explicit expression for the velocity $v = v(\partial_z P, c)$.

The heat flux accounts for mobile-phase convection and is expressed in terms of the enthalpy flux

$$E = \varepsilon \mathcal{H}(T, P, N), \quad (7)$$

neglecting gas- and solid-phase conduction. Here, first-order homogeneity is employed to express the enthalpy flux in terms of the enthalpy function and molar fluxes.

### Adsorption and heat transfer

The adsorption rate $R$ is defined by the stationary-phase mass balances

$$R = -\varphi\, \partial_t q, \quad (8)$$

where the volume ratio $\varphi$ ensures consistent mobile- and stationary-phase volume bases. The stationary-phase mass balances are modeled using the linear driving force (LDF) approximation

$$\partial_t q = \omega \odot (q^* - q), \quad (9)$$

with rate constants $\omega = [\omega_\alpha]_{\alpha \in \mathcal{C}}$ and equilibrium concentrations $q^* = [q^*_\alpha]_{\alpha \in \mathcal{C}}$. They are described by correlations, with parameters provided in the supplementary material.

### Adsorption equilibria

Following [2], we consider the multi-component Langmuir isotherm, which for a given component $\alpha \in \mathcal{C}$ is defined by

$$q^*_\alpha = q_{\max,\alpha} \frac{b_\alpha P_\alpha}{1 + \sum_{\beta \in \mathcal{C}} b_\beta P_\beta}. \quad (10)$$

Here, $q_{\max,\alpha}$ denotes the component's saturation capacity, $b_\alpha$ its Langmuir parameter, and $P_\alpha$ its partial pressure. The partial pressure is given by $P_\alpha = P c_\alpha / \sum_{\beta \in \mathcal{C}} c_\beta$. The remaining quantities are specified by temperature-dependent correlations

$$q_{\max,\alpha} = k_{1,\alpha} - k_{2,\alpha} T, \tag{11a}$$

$$b_\alpha = k_{3,\alpha} \exp\left(\frac{k_{4,\alpha}}{T}\right), \tag{11b}$$

with component-wise constants $k_{i,\alpha}$ for $i = 1,2,3,4$.

### Adsorption rates

For each component $\alpha \in \mathcal{C}$, the adsorption rate constant is defined as

$$\omega_\alpha = \Omega \frac{D_\alpha}{r_p^2}, \tag{12}$$

where $\Omega$ is a geometric factor, $D_\alpha$ denotes the component's effective pore diffusivity, and $r_p$ is the effective particle radius. Assuming effectively spherical particles, we take $\Omega = 15$. The apparent time constant $D_\alpha/r_p^2$ is modeled as a pressure-dependent quantity using the supercritical structural Langmuir model [2]

$$\frac{D_\alpha}{r_p^2} = k_\alpha \sqrt{P_{r,\alpha}} (1 + b_\alpha P_\alpha)^2, \tag{13}$$

where $k_\alpha$ is a kinetic parameter and $P_{r,\alpha} = P_\alpha/P_{c,\alpha}$ denotes the reduced pressure defined by the component's critical pressure $P_{c,\alpha}$. The Langmuir parameter $b_\alpha$ is consistent with the adsorption equilibrium model.

### Heat of adsorption

Neglecting wall effects, the heat transfer rate accounts for the heat of adsorption

$$Q = -(1-\varepsilon) \sum_{\alpha \in \mathcal{C}} \Delta H_\alpha \, \partial_t q_\alpha, \tag{14}$$

where $\Delta H = [\Delta H_\alpha]_{\alpha \in \mathcal{C}}$ denotes the component-wise heats of adsorption. Note that the heat of adsorption is taken to be negative for exothermic adsorption. Component-wise heats of adsorption are obtained from Clausius-Clapeyron analysis of the equilibrium isotherm [2].

## Thermodynamic constraints

The thermodynamic EOS is enforced through closures that recover the pseudo-homogeneous temperature and pressure profiles. We provide constraints on the internal energy and volume.

### Internal energy constraint

The total internal energy density is constrained by enforcing the decomposition into mobile- and stationary-phase contributions

$$u = \varepsilon u_{\text{gas}} + (1-\varepsilon) u_{\text{stationary}}. \tag{15}$$

Exploiting first-order homogeneity, the mobile-phase internal energy density is given by

$$u_{\text{gas}} = \mathcal{U}(T,P,c). \tag{16}$$

For the stationary phase, the internal energy density is split into its pore-phase and solid-phase contributions $u_{\text{stationary}} = u_{\text{pore}} + u_{\text{solid}}$. The pore-phase internal energy density is similarly given by

$$u_{\text{pore}} = \mathcal{U}(T,P,q). \tag{17}$$

Moreover, the solid-phase internal energy density is defined as

$$u_{\text{solid}} = \rho_{\text{solid}} c_{p,\text{solid}} (T - T_{\text{ref}}), \tag{18}$$

where $\rho_{\text{solid}}$ denotes the solid-phase density, $c_{p,\text{solid}}$ the heat capacity, and $T_{\text{ref}}$ a reference temperature.

### Volume constraint

The total gas volume comprises the mobile-phase and pore gas volumes. Due to sorption effects, the pore gas volume is not well-described by the EOS. Instead, a volume constraint is enforced on the mobile-phase gas

$$\mathcal{V}(T,P,c) = 1, \tag{19}$$

which, by first-order homogeneity, ensures that the relation $\mathcal{V}(T,P,n) = V_{\text{gas}}$ holds.

## Initial and boundary conditions

The differential mass and energy balances are closed by a set of initial and boundary conditions.

### Initial conditions

Initial conditions are prescribed for the differential states $(c, q, u)$ at time $t = 0$ across the axial domain $z \in \Omega$

$$c(0,z) = c_0(z), \tag{20a}$$

$$q(0,z) = q_0(z), \tag{20b}$$

$$u(0,z) = u_0(z). \tag{20c}$$

From these, consistent initial profiles of the algebraic states $(T, P)$ are subsequently obtained from the thermodynamic constraints (15) and (19).

### Boundary conditions

Boundary conditions are prescribed directly on the boundary fluxes

$$N(t,0) = N_{\text{in}}(t), \tag{21a}$$

$$E(t,0) = E_{\text{in}}(t), \tag{21b}$$

$$N(t,L) = N_{\text{out}}(t), \tag{21c}$$

$$E(t,L) = E_{\text{out}}(t). \tag{21d}$$

We employ flow-driven boundary conditions, specifying the upstream flows and downstream pressure. Consequently, the inlet fluxes are defined by

$$N_{\text{in}}(t) = \frac{f_{\text{in}}(t)}{S_{\text{gas}}}, \tag{22a}$$

$$E_{\text{in}}(t) = \varepsilon \frac{h_{\text{in}}(t)}{S_{\text{gas}}}, \tag{22b}$$

where $f_{\text{in}} = [f_{\text{in},\alpha}]_{\alpha \in \mathcal{C}}$ denotes the inlet molar flow rates and $h_{\text{in}} = \mathcal{H}(T_{\text{in}}, P_{\text{in}}, f_{\text{in}})$, the corresponding inlet enthalpy flow, specified by the inlet temperature $T_{\text{in}}$ and pressure $P_{\text{in}}$. Moreover, $S_{\text{gas}} = V_{\text{gas}}/L$ denotes the cross-sectional gas area. At the outlet, the prescribed pressure $P_{\text{out}}(t)$ determines the outlet pressure gradient $\partial_z P(t,L)$ and thus the outlet velocity $v_{\text{out}}(t)$ via the momentum balance (5). To prevent backflow, we adopt a check-valve formulation and determine

$$v_{\text{out}}(t) = \max_\delta\{0, v(\partial_z P(t,L), c(t,L))\}, \quad (23)$$

where the smoothed maximum operator is given by $\max_\delta\{a,b\} = \left(a + b + \sqrt{(a-b)^2 + \delta}\right)/2$ with $\delta = 10^{-4}$. The outlet fluxes are then defined by

$$N_{\text{out}}(t) = v_{\text{out}}(t)c(t,L), \quad (24a)$$

$$E_{\text{out}}(t) = \varepsilon \mathcal{H}\big(T(t,L), P(t,L), N_{\text{out}}(t)\big). \quad (24b)$$

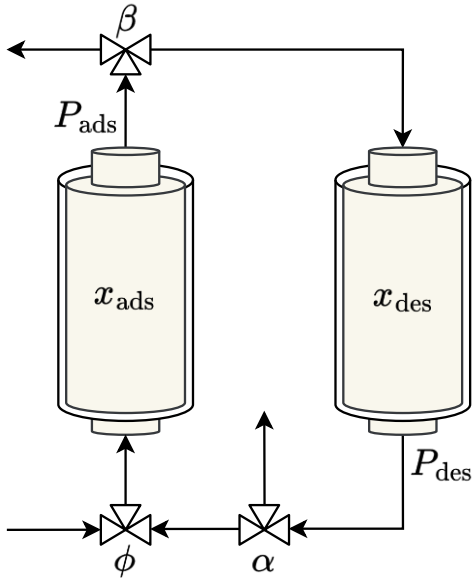

**Figure 1.** Diagram of the two-bed PSA superstructure.

## PSA SUPERSTRUCTURE

The PSA process is simulated by periodically sequencing a set of operating steps (pressurization, adsorption, blowdown, desorption, purge, and pressure equalization) across multiple adsorption beds. In a conventional two-bed PSA configuration, the operating steps are symmetric and may be divided into two complementary half-cycles: an adsorption half-cycle and a desorption half-cycle. While one bed undergoes the adsorption half-cycle, the other simultaneously undergoes the desorption half-cycle. Under this symmetry, it is sufficient to simulate the two-bed interactions across a single half-cycle, with the remaining half-cycle obtained by swapping beds. This motivates the PSA superstructure in [3], as shown in Figure 1. The superstructure defines the interconnections between the adsorption beds through idealized valves and imposed adsorption and desorption pressure levels. We omit the modeling of auxiliary equipment and set pressure levels via boundary conditions.

## Mathematical formulation

We formulate the PSA superstructure over a half-cycle time horizon $t \in [0, \tau]$. Each bed is represented by an instance of the adsorption column model described previously. For each bed, we denote its differential states by $x = [c; q; u]$ and its algebraic states by $y = [T, P]$. The two beds are distinguished by subscripts indicating whether the bed undergoes adsorption or desorption. The left-most bed in Figure 1 undergoes the adsorption half-cycle, while the right-most bed undergoes the desorption half-cycle.

### Boundary conditions

In the left-most bed, a fraction $\phi(t)$ of the constant feed stream $(f_{\text{in}}, h_{\text{in}})$ enters from the bottom inlet, together with a fraction $\alpha(t)$ of the right-most bed outlet

$$N_{\text{in,ads}}(t) = \phi(t)\frac{f_{\text{in}}}{S_{\text{gas}}} + \alpha(t)N_{\text{out,des}}(t), \quad (25a)$$

$$E_{\text{in,ads}}(t) = \varepsilon\phi(t)\frac{h_{\text{in}}}{S_{\text{gas}}} + \alpha(t)E_{\text{out,des}}(t). \quad (25b)$$

Similarly, a fraction $\beta(t)$ of the left-most bed outlet enters the top inlet of the right-most bed

$$N_{\text{in,des}}(t) = \beta(t)N_{\text{out,ads}}(t), \quad (26a)$$

$$E_{\text{in,des}}(t) = \beta(t)E_{\text{out,ads}}(t). \quad (26b)$$

Here, the outlet fluxes are defined by (24), via the prescribed adsorption pressure $P_{\text{ads}}(t)$ and desorption pressure $P_{\text{des}}(t)$. The superstructure enforces unidirectional flow through the check-valve formulation (23), allowing the half-cycle to be simulated without flow reversal.

### Cyclic steady state

A consistent PSA cycle is obtained from the simulated half-cycle at CSS. Exploiting the symmetry of the two-bed superstructure, the full PSA cycle is recovered by imposing periodic boundary conditions, such that the terminal state of one bed at the end of a half-cycle provides the initial condition for the other bed at the beginning of the subsequent half-cycle, and vice versa. Accordingly, the differential states of the superstructure must solve the boundary value problem (BVP)

$$x_{\text{des}}(\tau, L-z) = x_{\text{ads}}(0, z), \quad (27a)$$

$$x_{\text{ads}}(\tau, L-z) = x_{\text{des}}(0, z), \quad (27b)$$

for all $z \in \Omega$. Here the axial reversal reflects the flow configuration of the superstructure. Consistent periodic profiles of the algebraic states are recovered from the thermodynamic constraints as part of the CSS solution.

## NUMERICAL SIMULATION

We simulate the PSA superstructure by discretizing the governing system of PDAEs in space using the method of lines. The resulting semi-discrete system of DAEs is then integrated in time.

### Finite volume discretization

We discretize the axial domain $\Omega$ into $K$ uniform volumes $\Omega_h^k = \left[z_h^{k-1/2}, z_h^{k+1/2}\right]$ of length $h = L/K$, and consider a cell-centered finite volume approximation $w_h = [w_h^k]_{k=1}^K$ of a given quantity $w$. For a volume $\Omega_h^k$, we define its midpoint $z_h^k = (k - 1/2)h$ and interfaces $z_h^{k\pm 1/2} = z_h^k \pm h/2$, and approximate the volume-averaged quantity

$$w_h^k(t) \approx \langle w \rangle_h^k(t) = \frac{1}{h}\int_{z_h^{k-1/2}}^{z_h^{k+1/2}} w(t,z)\,dz, \tag{28}$$

by midpoint evaluation $w_h^k(t) = w(t, z_h^k)$. Applying the fundamental theorem of calculus and the same midpoint approximation to the volume-averaged mass and energy balances (1), we find that

$$\partial_t c_h^k = -\frac{1}{h}\left(N_h^{k+1/2} - N_h^{k-1/2}\right) + R_h^k, \tag{29a}$$

$$\partial_t u_h^k = -\frac{1}{h}\left(E_h^{k+1/2} - E_h^{k-1/2}\right) + Q_h^k, \tag{29b}$$

where $N_h^{k\pm 1/2}$ and $E_h^{k\pm 1/2}$ denote the interface molar and heat fluxes, respectively, and where $R_h^k$ and $Q_h^k$ denote the midpoint approximations of (8) and (14)

$$R_h^k = -\varphi\,\partial_t q_h^k, \tag{30a}$$

$$Q_h^k = -(1-\varepsilon)\sum_{\alpha \in \mathcal{C}} \Delta H_\alpha (\partial_t q_\alpha)_h^k. \tag{30b}$$

Here, $\partial_t q_h^k = [(\partial_t q_\alpha)_h^k]_{\alpha \in \mathcal{C}}$ is defined by the midpoint approximations of the LDF approximation

$$\partial_t q_h^k = \omega_h^k \odot ((q^*)_h^k - q_h^k), \tag{31}$$

with $(q^*)_h^k = q^*(T_h^k, P_h^k, c_h^k)$ and $\omega_h^k = \omega(T_h^k, P_h^k, c_h^k)$.

### Numerical fluxes

Since the superstructure enforces unidirectional flow, stability is ensured by prescribing upwind numerical fluxes at the volume interfaces. They are defined as

$$N_h^{k+1/2} = v\left(\partial_z P_h^{k+1/2}, c_h^k\right) c_h^k, \tag{32a}$$

$$E_h^{k+1/2} = \varepsilon \mathcal{H}\left(T_h^k, P_h^k, N_h^{k+1/2}\right), \tag{32b}$$

where the interface pressure gradient is approximated by $\partial_z P_h^{k+1/2} = (P_h^{k+1} - P_h^k)/h$. This results in a monotone first-order accurate spatial discretization.

### Thermodynamic constraints

The thermodynamic constraints are imposed locally in each finite volume $\Omega_h^k$. Applying the midpoint approximation to the internal energy constraint (15) yields

$$u_h^k = \varepsilon (u_{\text{gas}})_h^k + (1-\varepsilon)(u_{\text{stationary}})_h^k, \tag{33}$$

with $(u_{\text{stationary}})_h^k = (u_{\text{solid}})_h^k + (u_{\text{pore}})_h^k$. Here, the corresponding phase contributions are given by midpoint evaluation of (16)-(18) as

$$(u_{\text{gas}})_h^k = \mathcal{U}(T_h^k, P_h^k, c_h^k), \tag{34a}$$

$$(u_{\text{pore}})_h^k = \mathcal{U}(T_h^k, P_h^k, q_h^k), \tag{34b}$$

$$(u_{\text{solid}})_h^k = \rho_{\text{solid}} c_{p,\text{solid}} (T_h^k - T_{\text{ref}}). \tag{34c}$$

Similarly, the volume constraint (19) is also enforced by midpoint evaluation

$$\mathcal{V}(T_h^k, P_h^k, c_h^k) = 1. \tag{35}$$

### Initial and boundary conditions

The initial conditions are also discretized by midpoint evaluation

$$c_h^k(0) = c_0(z_h^k), \tag{36a}$$

$$q_h^k(0) = q_0(z_h^k), \tag{36b}$$

$$u_h^k(0) = u_0(z_h^k), \tag{36c}$$

from which the initial algebraic states are recovered by the discrete thermodynamic constraints. The inlet boundary conditions are simply given by the inlet fluxes

$$N_h^{1/2}(t) = N_{\text{in}}(t), \tag{37a}$$

$$E_h^{1/2}(t) = E_{\text{in}}(t), \tag{37b}$$

and the outlet boundary conditions are given by the upwind fluxes

$$N_h^{K+1/2}(t) = v\left(\partial_z P_h^{K+1/2}(t), c_h^K(t)\right) c_h^K(t), \tag{38a}$$

$$E_h^{K+1/2}(t) = \varepsilon \mathcal{H}\left(T_h^K(t), P_h^K(t), N_h^{K+1/2}(t)\right), \tag{38b}$$

Here, the outlet pressure gradient is approximated by $\partial_z P_h^{K+1/2}(t) = \left(P_{\text{out}}(t) - P_h^K(t)\right)/(h/2)$ using the prescribed outlet pressure $P_{\text{out}}(t)$.

### Semi-discrete system

We let $\mathbf{x}_h = [c_h; q_h; u_h]$ and $\mathbf{y}_h = [T_h; P_h]$ denote the finite-volume discretizations of the differential and algebraic states, respectively. For the two-bed PSA superstructure, the corresponding discretized states of both beds are collected into global differential and algebraic state vectors $\mathbf{x}$ and $\mathbf{y}$. The spatial discretization of the superstructure then yields a semi-discrete system of semi-explicit index-1 DAEs

$$\dot{\mathbf{x}} = f(\mathbf{x}, \mathbf{y}, p), \quad \mathbf{x}(0) = \mathbf{x}_0, \tag{39a}$$

$$0 = g(\mathbf{x}, \mathbf{y}, p), \tag{39b}$$

where $\mathbf{x}_0$ is defined by the discretized initial conditions, and $p$ the prescribed boundary conditions.

### Initialization

Before integration, we solve for consistent initial algebraic states $\mathbf{y}(0) = \mathbf{y}_0$ satisfying

$$g(\mathbf{x}_0, \mathbf{y}_0, p) = 0. \quad (40)$$

To this end, we employ Newton's method with backtracking line search, which restricts trial iterates to the feasible domain of the model functions and prevents evaluation at physically inadmissible states.

### Diagonally implicit Runge-Kutta methods

Time integration of (39) is performed using diagonally implicit Runge-Kutta (DIRK) methods [5]. Given states $(\mathbf{x}_n, \mathbf{y}_n)$ at time $t = t_n$, an $s$-stage DIRK method advances to the states $(\mathbf{x}_{n+1}, \mathbf{y}_{n+1})$ at time $t = t_{n+1}$, with step size $h_n = t_{n+1} - t_n$, by computing stage values $(\mathbf{X}_{n,i}, \mathbf{Y}_{n,i})$ satisfying the implicit stage equations

$$\mathbf{X}_{n,i} = \mathbf{x}_n + h_n \sum_{j=1}^{s} a_{i,j} f(\mathbf{X}_{n,j}, \mathbf{Y}_{n,j}, p), \quad (41a)$$

$$0 = g(\mathbf{X}_{n,i}, \mathbf{Y}_{n,i}, p), \quad (41b)$$

defined at stage times $T_{n,i} = t_n + c_i h$, for $i = 1, \dots, s$. The advancing states are found via the stage values using the quadrature rule

$$\mathbf{x}_{n+1} = \mathbf{x}_n + h_n \sum_{i=1}^{s} b_i f(\mathbf{X}_{n,i}, \mathbf{Y}_{n,i}, p), \quad (42a)$$

$$0 = g(\mathbf{x}_{n+1}, \mathbf{y}_{n+1}). \quad (42b)$$

The DIRK method is characterized by a Butcher tableau with coefficients $a = [a_{i,j}]_{i,j=1}^{s}$, $b = [b_i]_{j=1}^{s}$, and $c = [c_i]_{i=1}^{s}$, where $a_{i,j} = 0$ for $i < j$. In this work, we employ singly-diagonally implicit Runge-Kutta (SDIRK) methods, and in particular explicit first-stage SDIRK (ESDIRK) methods, with constant diagonal coefficients $a_{i,i} = \gamma$ and $a_{1,1} = 0$. We further adopt stiffly accurate schemes where $b_i = a_{s,i}$, such that the final stage directly gives the advancing state. This combination of properties renders stiffly accurate ESDIRK methods particularly efficient. The lower-triangular structure enables sequential solution of the stage equations via Newton's method, while the constant diagonal allows reuse of a single Jacobian factorization across all stages within a time step, making quasi-Newton schemes especially effective. The remaining coefficients are then chosen to ensure high-order accuracy and stability. Step sizes are chosen adaptively, to control the temporal error and to handle transient stiffness arising from changes in boundary conditions. We use PID-based step-size control via embedded error estimates, as described in [5].

## CYCLIC STEADY-STATE SIMULATION

While the previous section describes numerical simulation of a single half-cycle of the PSA superstructure, a physically consistent PSA cycle is obtained only when the half-cycle satisfies the CSS conditions defined by the BVP (27). Using the method-of-lines discretization, the CSS BVP is solved numerically using shooting methods.

### Single shooting

Single shooting aims to solve the BVP by iteratively integrating the semi-discrete DAE system and updating the initial state until CSS is achieved. Using the ESDIRK integrators, we introduce the half-cycle propagator $\Phi$, which, for given initial differential states $\mathbf{x}_0$ and boundary conditions $p$, returns the terminal states $\Phi(\mathbf{x}_0, p)$ at the end of the half cycle via numerical integration over $t \in [0, \tau]$. Additionally, we define the operator $\Pi$, which implements the permutation of the superstructure states that swaps the two beds and applies the axial reversal. The discretized CSS BVP reduces to the fixed-point problem

$$\Pi \Phi(\mathbf{x}, p) = \mathbf{x}, \quad (43)$$

whose solution defines initial states for a consistent PSA cycle. Alternatively, we adopt the residual formulation

$$R(\mathbf{x}) = \Pi \Phi(\mathbf{x}, p) - \mathbf{x}, \quad (44)$$

and seek a solution to $R(\mathbf{x}) = 0$. The fixed-point formulation of single shooting lends itself to simple fixed-point (Picard) iterations and acceleration techniques, such as Anderson acceleration (AA). In contrast, Newton's method may be used to solve the residual formulation.

## IMPLEMENTATION DETAILS

The simulation framework is implemented in the Julia 1.10 language and executed on an Intel Xeon E5-2650 (2.2 GHz) workstation. Model derivatives are computed by combining analytical derivatives of the finite-volume discretization with forward-mode automatic differentiation of the thermodynamic functions provided by `Clapeyron.jl` [6]. Dense linear algebra is performed with Intel MKL-accelerated BLAS, while sparse model Jacobians are assembled explicitly and factorized using sparse direct linear algebra within the Newton schemes used for consistent initialization and ESDIRK time integration. Sensitivities of the half-cycle propagators are computed via staggered direct forward sensitivity analysis of the ESDIRK method and are used in the Newton-based single shooting solver.

**Table 1.** Process conditions of the 4-step PSA half-cycle.

| Step | PR | ADS | PUR | EQ |
|---|---|---|---|---|
| Time [s] | 20 | 20 | 45 | 5 |
| $P_{ads}$ [bar] | 8 | 8 | 8 | 4.5 |
| $P_{des}$ [bar] | 1 | 1 | 1 | 8 |
| α [-] | 0 | 0 | 0 | 0 |
| β [-] | 0 | 0 | 0.5 | 1 |
| φ [-] | 1 | 1 | 1 | 0 |

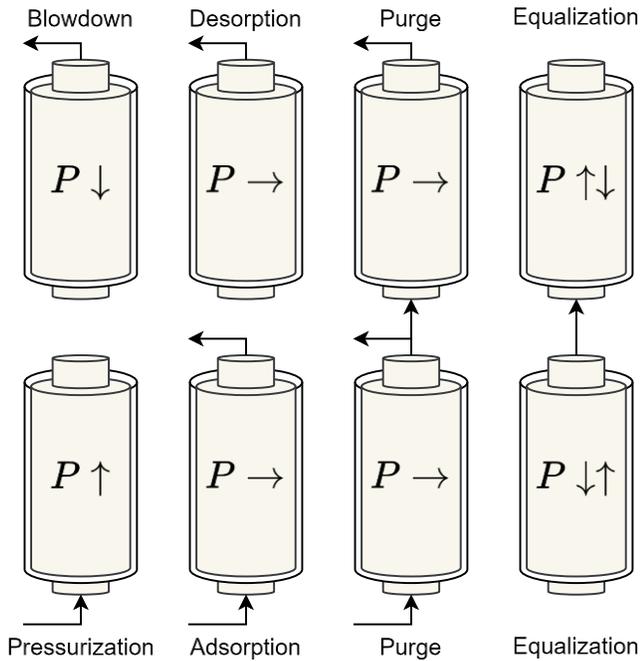

**Figure 2.** Diagram of the 8-step PSA cycle.

**Table 2.** Column and feed parameters.

| L [m] | D [m] | $T_{in}$ [K] | $P_{in}$ [bar] | $v_{in}$ [cm/s] |
|---|---|---|---|---|
| 2 | 1 | 298 | 8 | 5 |

**Table 3.** Simulated cycle performance (Tol=$10^{-8}$).

| Statistic | EOS | K=10 | K=20 | K=40 | K=80 |
|---|---|---|---|---|---|
| Purity [%] | Ideal | 95.4% | 95.8% | 96.0% | 96.2% |
|  | Real | 95.3% | 95.8% | 96.1% | 96.1% |
| Recov. [%] | Ideal | 45.9% | 45.9% | 45.9% | 45.9% |
|  | Real | 45.7% | 45.7% | 45.7% | 45.8% |

## NUMERICAL EXPERIMENTS

We consider simulation of the example two-bed PSA cycle described in Table 1 and shown in Figure 2. The columns are fed dry air at ambient temperature, supplied at constant pressure and flow rate. Column and feed parameters are summarized in Table 2. Studying a converged CSS, we first assess how spatial resolution and thermodynamics affect simulated cycle performance.

**Table 5.** Solver statistics (EOS=Real, K=40, Tol=$10^{-6}$).

| Statistic | **Picard** | **AA** | **Newton** |
|---|---|---|---|
| Iterations [#] | 1123 | 416 | **3** |
| Time [hr:min] | 01:19 | 00:23 | **00:02** |

Using tight integration tolerances, purity and recovery are compared for ideal and real-fluid thermodynamics across increasing finite-volume resolutions. As shown in Table 3, the predicted product purity increases with spatial resolution, whereas the recovery remains essentially unchanged. Differences between ideal and real-fluid thermodynamics are negligible, on the order of 1–2‰, indicating that spatial discretization effects dominate.

The dominant computational task is governed by time integration, and in particular thermodynamic property evaluation. Table 4 compares several ESDIRK schemes from [5] for CSS simulation across error tolerances. While higher-order methods require fewer time steps, their larger stage counts make lower-order schemes slightly faster at loose tolerances. However, higher-order schemes are more efficient at stricter tolerances. Simulation cost scales linearly in $K$, as the runtime is dominated by model evaluation rather than the efficient sparse linear algebra. Since model evaluation is dominated by thermodynamics, ideal conditions are up to 50% faster to simulate across spatial resolutions.

Beyond the cost of individual cycle simulations, the computation of the CSS hinges on how efficiently the CSS BVP is solved. Using the 4$^{th}$ order ESDIRK method, Table 5 compares Picard iterations, Anderson acceleration, and Newton-based solution. Newton's method converges in only three iterations with quadratic convergence, achieving a tighter solution tolerance in minutes, whereas Picard and Anderson-accelerated iterations require hundreds of iterations and substantially longer runtimes. The dominant eigenvalue of the propagation operator lies within the integration tolerance of unity, making fixed-point convergence noisy, necessitating tight integration tolerances.

Finally, the two half-cycles are stitched together to recover the full periodic operation and simulated spatio-temporal composition profiles are shown in Figure 3. The resulting product flow rate of approximately 450 [m$^3$/h] at ambient conditions falls well within the range of small-scale PSA-based air separation units.

**Table 4:** CSS simulation statistics (EOS=Real, K=80).

| Method | Statistic | Tol=$10^{-2}$ | Tol=$10^{-4}$ | Tol=$10^{-6}$ | Tol=$10^{-8}$ |
|---|---|---|---|---|---|
| **ESDIRK2(1)3L[2]SA** | Steps [#] | 178 | 266 | 1589 | 14542 |
| 2$^{nd}$ order 3 stages | Time [min:sec] | **00:04** | **00:06** | 00:31 | 04:43 |
| **ESDIRK3(2)5L[2]SA** | Steps [#] | 143 | 163 | 282 | 725 |
| 3$^{rd}$ order 5 stages | Time [min:sec] | 00:05 | **00:06** | **00:10** | 00:27 |
| **ESDIRK4(3)7L[2]SA** | Steps [#] | **129** | **144** | **233** | **377** |
| 4$^{th}$ order 7 stages | Time [min:sec] | 00:06 | 00:07 | 00:11 | **00:20** |

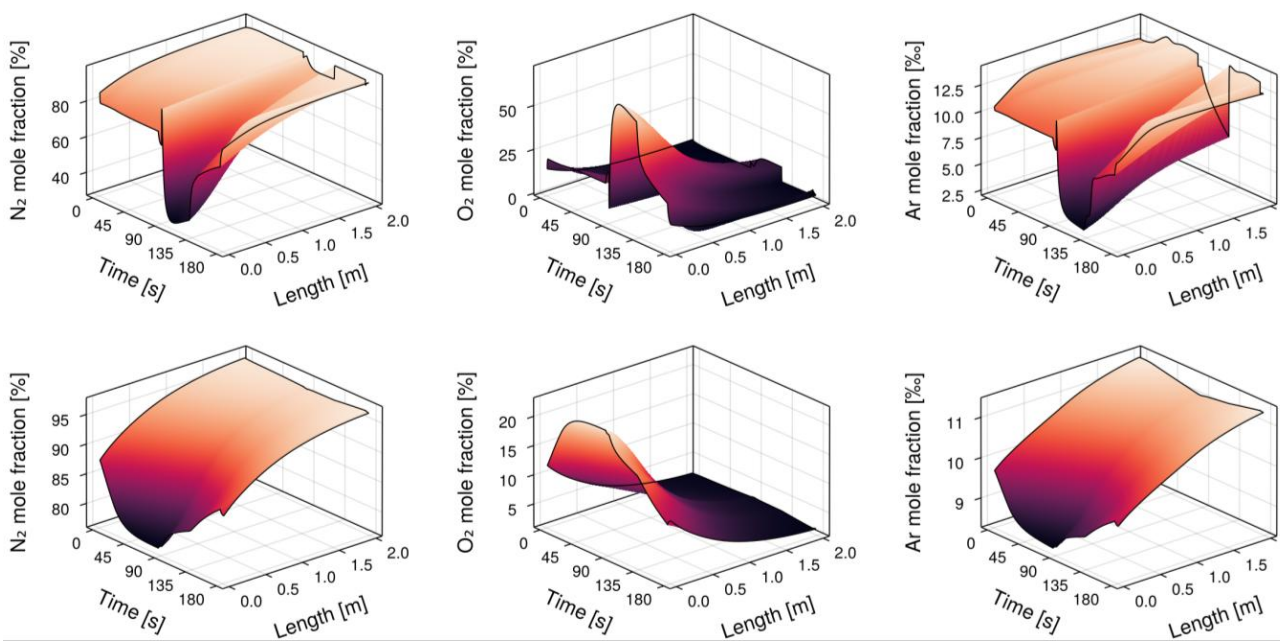

**Figure 3:** Fraction profiles throughout the 8-step PSA cycle at CSS (top: mobile, bottom: stationary).

## CONCLUSION

This work presented a first-principles, dynamic modeling framework for PSA-based nitrogen generation. Thermodynamically consistent column models, combined with finite-volume discretization and implicit time integration, enable systematic simulation and CSS computation via shooting methods. Numerical experiments show that solver choice strongly affects computational efficiency, with Newton-based methods converging substantially faster than fixed-point approaches. The framework provides a systematic basis for optimization of PSA cycles for nitrogen generation, following [3].

## DIGITAL SUPPLEMENTARY MATERIAL

Adsorbent and adsorption rate parameters are available in the LAPSE at LAPSE:2026.0016.

## ACKNOWLEDGEMENTS

This project has been funded by the MissionGreenFuels project DynFlex under The Innovation Fund Denmark project no. 1150-00001B.

## AUTHOR IDENTIFIERS

Author ORCIDs:
Schytt MJ: 0009-0001-8143-1878
Biegler LT: 0000-0003-3875-4441
Jørgensen JB: 0000-0001-9799-2808